\documentclass[iop]{emulateapj}

\usepackage{soul}

\newcommand{\czw}{\ensuremath{^{12}\mem{C}}}
\newcommand{\mem}[1]{\ensuremath{\mathrm{ #1}}}
\newcommand{\p}{\ensuremath{\mem{p}}}
\newcommand{\ndr}{\ensuremath{^{13}\mem{N}}}
\newcommand{\natlog}[2]{\ensuremath{#1\times 10^{#2}}}
\newcommand{\kelv}{\ensuremath{\,\mathrm K}}
\newcommand{\kap}[1]{Sect.\,\ref{#1}}
\newcommand{\lsun}{\ensuremath{\, {\rm L}_\odot}}
\newcommand{\msun}{\ensuremath{\, {\rm M}_\odot}}
\newcommand{\abb}[1]{Fig.\,\ref{#1}}

\newcommand{\jahre}{\ensuremath{\, \mathrm{yr}}}
\newcommand{\apgeq}{\ensuremath{\stackrel{>}{_\sim}}}
\newcommand{\apleq}{\ensuremath{\stackrel{<}{_\sim}}}

\slugcomment{Submitted: October 19, 2013; accepted July 17, 2014}

 \usepackage{natbib}

\shorttitle{3-D H-ingestion flash simulations} \shortauthors{Herwig
etal.}

\begin{document}


\title{Global non-spherical oscillations in 3-D $4\pi$ simulations of
  the H-ingestion flash}

\author{
Falk Herwig\altaffilmark{1,2} 
Paul R. Woodward\altaffilmark{3},
Pei-Hung Lin\altaffilmark{3}, 
Mike Knox\altaffilmark{3},
Chris Fryer\altaffilmark{4}
}
\altaffiltext{1}{Department of Physics \& Astronomy, University of
Victoria, Victoria, BC V8P5C2, Canada}
\altaffiltext{2}{Turbulence in Stellar Astrophysics Program, New Mexico
  Consortium, Los Alamos, NM 87544, USA}
 \altaffiltext{3}{LCSE \&
Department of Astronomy, University of Minnesota, Minneapolis, MN 55455,
USA} \altaffiltext{4}{Computational Computer Science Division, Los
Alamos National Laboratory, Los Alamos, NM 87545, USA; Physics
Department, University of Arizona, Tucson, AZ 85721, USA}
\email{fherwig@uvic.ca}

\begin{abstract}
We performed 3-D simulations of proton-rich material entrainment into
\czw-rich He-shell flash convection and the subsequent H-ingestion
flash that took place in the post-AGB star Sakurai's
object. Observations of the transient nature and anomalous abundance
features are available to validate our method and assumptions, with
the aim to apply them to very low metallicity stars in the future. We
include nuclear energy feedback from H burning and cover the full
$4\pi$ geometry of the shell. Runs on $768^3$ and $1536^3$ grids agree
well with each other and have been followed for $1500\mathrm{min}$ and
$1200\mathrm{min}$. After a $850\mem{min}$ long quiescent entrainment
phase the simulations enter into a global non-spherical oscillation
that is launched and sustained by individual ignition events of H-rich
fluid pockets. Fast circumferential flows collide at the antipode and
cause the formation and localized ignition of the next H-overabundant
pocket. The cycle repeats for more than a dozen times while its
amplitude decreases. During the global oscillation the entrainment
rate increases temporarily by a factor $\approx 100$. Entrained
entropy quenches convective motions in the upper layer until the
burning of entrained H establishes a separate convection zone. The
lower-resolution run hints at the possibility that another global
oscillation, perhaps even more violent will follow. The location of
the H-burning convection zone agrees with a 1-D model in which the
mixing efficiency is calibrated to reproduce the light curve. The
simulations have been performed at the NSF Blue Waters supercomputer
at NCSA.
 \end{abstract}

%

\section{Introduction} 
\label{sec:intro} 

Convective-reactive H-combustion events are encountered when the
Damk\"ohler number $Da = \tau_\mem{conv/adv} / \tau_{\czw}(p) \sim 1$,
where $\tau_{\czw}(p)$ is the reaction time scale for the
$\czw(\p,\gamma)\ndr$ reaction and $\tau_\mem{conv/adv}$ is the
convective advection time scale \citep{dimotakis:05}.  In such events
H is ingested into He-shell flash convection, and this is encountered
in many instances of stellar evolution, especially at very low or zero
metallicity. Examples  have been found in 1-D stellar evolution
  simulations and include the He-core and He-shell flashes
with H-ingestion events in low-mass stars with [Fe/H]$\leq -2$
\citep[e.g.][]{fujimoto:00,iwamoto:04,campbell:08,Cristallo:2009cu,Lau:2009bd,campbell:10},
and in massive stars \citep{Ekstrom:2008fq,heger:10}.

During H-ingestion events, nuclear energy release on the convective
turn-over time scale is coupled with multi-scale turbulent mixing, and
therefore results from 1-D simulations are unreliable
\citep{herwig:10a,Arnett:2013wi}.  \citet{mocak:11a} presented
simulations of a H-burning convective shell induced by an artificially
inward shifted H-profile on top of the He-core flash convection
zone. However, in their H-burning convection region the nuclear time
scale of the $\czw(p,\gamma)\ndr$ reaction can be estimated from the
available $T$ and $\rho$ information to be
$\apgeq\natlog{4}{4}\mathrm{s}$, which implies
$\mathrm{Da}\apleq0.01$. Therefore, their H-burning convection zone is not
convective-reactive, and therefore not a H-ingestion event in terms of
our definition. There is a small amount of H penetrating through the
interface between the two convection zones, and it does not have
dynamical relevance (just as in our simulations in the first few
hours, see below).

\citet{stancliffe:11} present H-ingestion AGB shell flash convection
simulations for a low-metallicity case.  They report a H-burning
luminosity difference of orders of magnitude between simulations with
two different grid resolutions, and an almost complete transport of
protons through the entire convection zone all the way to the bottom, 
where the highest temperatures ($T_\mem{max}\sim
\natlog{2.3}{8}\kelv$) are reached and the protons are reacting with
\czw. Contrary to one-dimensional stellar evolution calculations,
which place the location of H-burning in the He-shell flash convection
zone at $T_\mem{max}\sim 1.0 - \natlog{1.5}{8}\kelv$, those 3-D
hydrodynamic simulations do not show any apparent hydrodynamic
feedback from the energy release from H-burning. 

\begin{figure}[t]
 \includegraphics[width=0.5\textwidth]{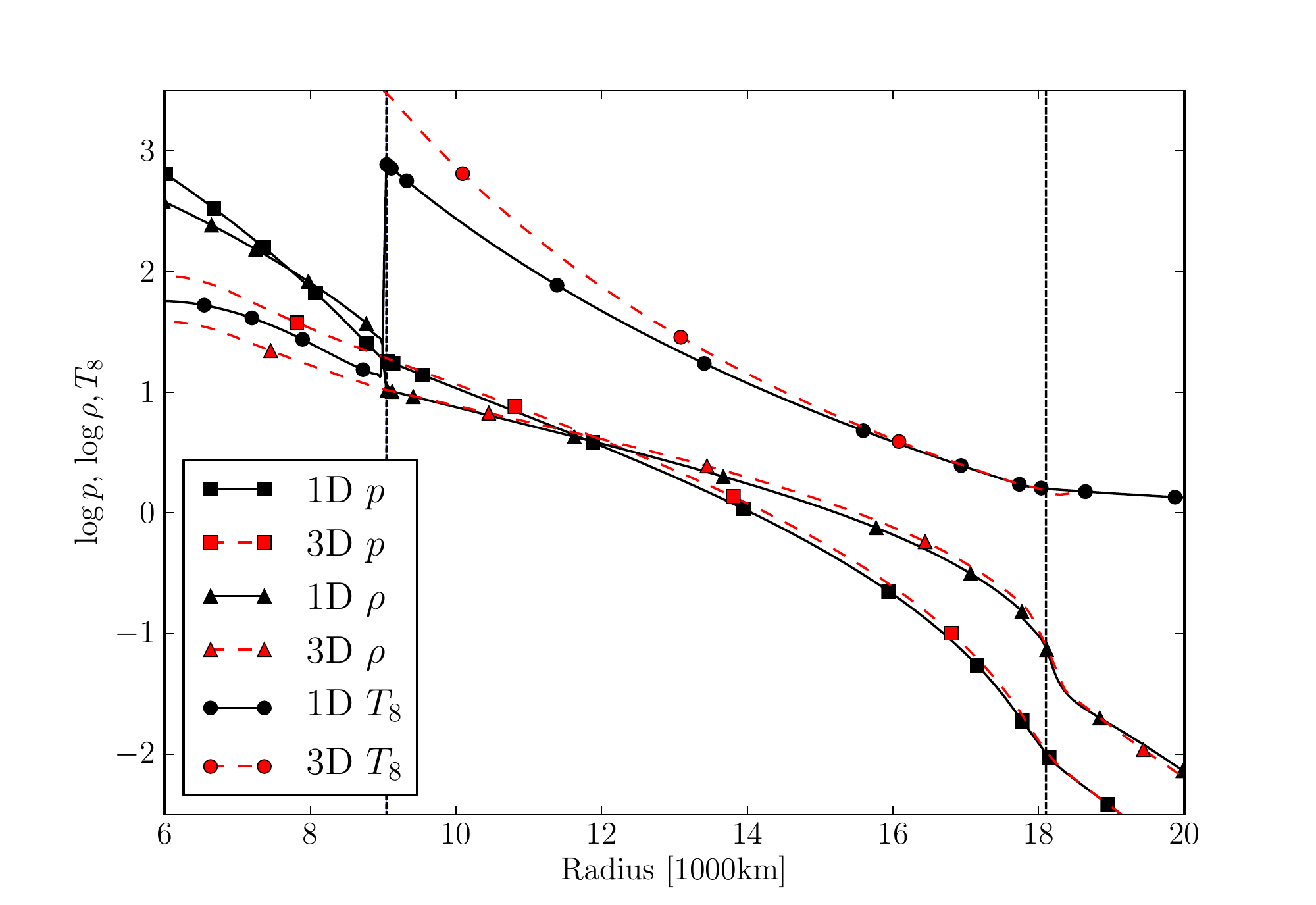} 
 \includegraphics[width=0.5\textwidth]{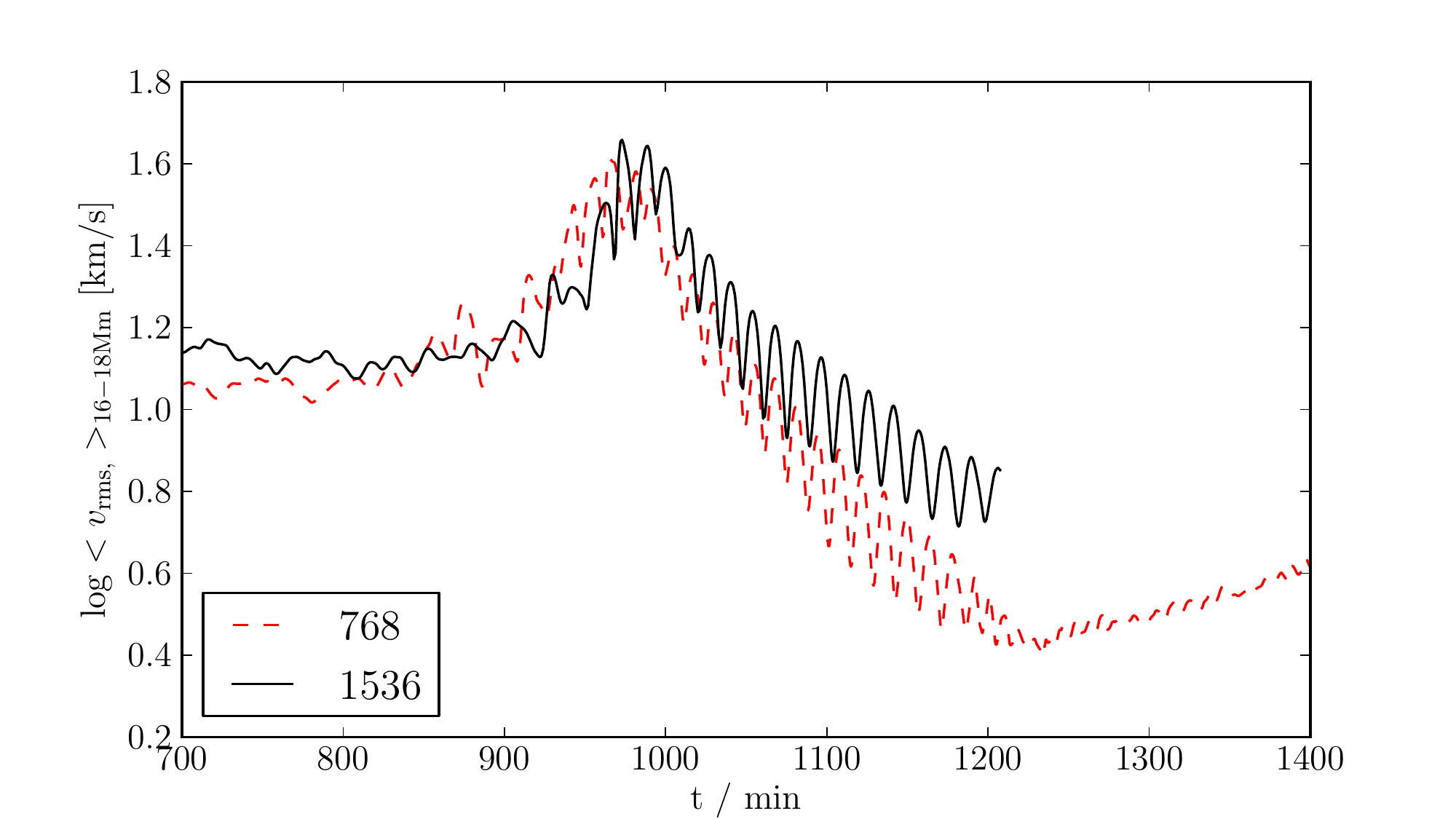} 
 \caption{Top: Initial piecewise polytropic setup of 3-D simulation in
   comparison with the 1-D stellar evolution model for Sakurai's
   object at time $t_0$ \citep[Fig.\,2][]{herwig:10a}. Shown are
   pressure $p$ and density $\rho$ in code units ($[p]=10^{19}
   \mathrm{g\,cm^{-1}\, s^{-2}}$), $[\rho] = 10^3
   \mathrm{g\,cm^{-3}}$), and temperature ($[T]=10^8\kelv$).  Vertical
   dotted lines indicate the location of the convection
   boundary. Bottom: Spherically averaged rms velocity averaged over
   the top region of the convection zone from $16$ to
   $\natlog{18}{3}\mem{km}$. The signature of a global oscillation is
   evident in these averaged velocities.}
\label{fig:initial_star} 
\end{figure}




\begin{figure*}
\includegraphics[width=0.333\textwidth]{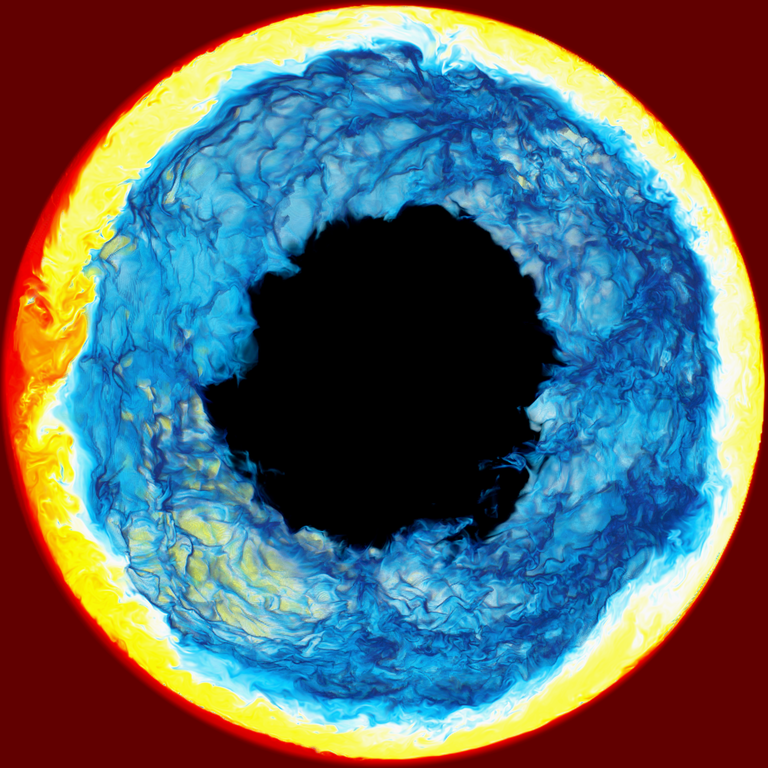}
\includegraphics[width=0.333\textwidth]{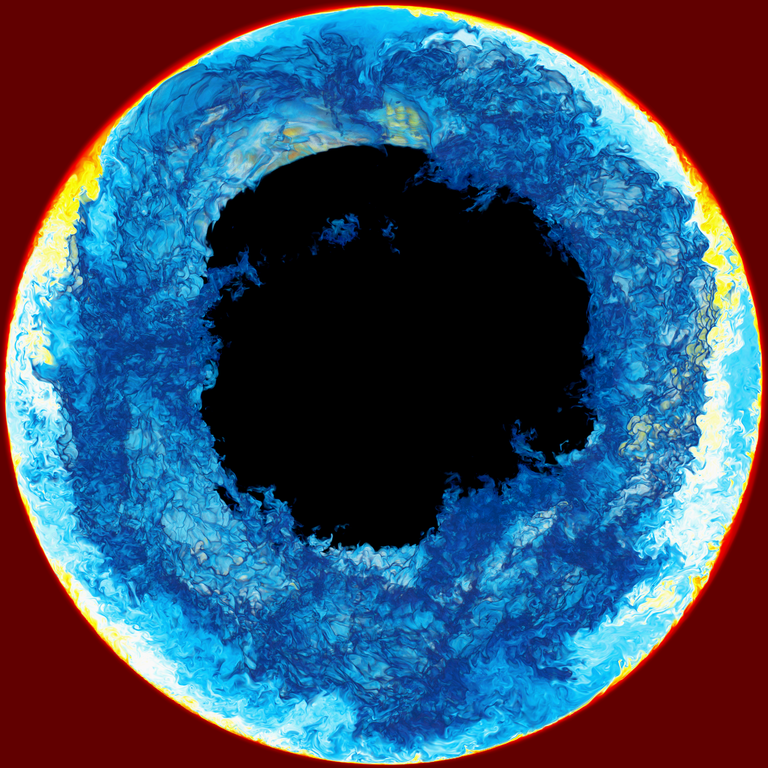}
\includegraphics[width=0.333\textwidth]{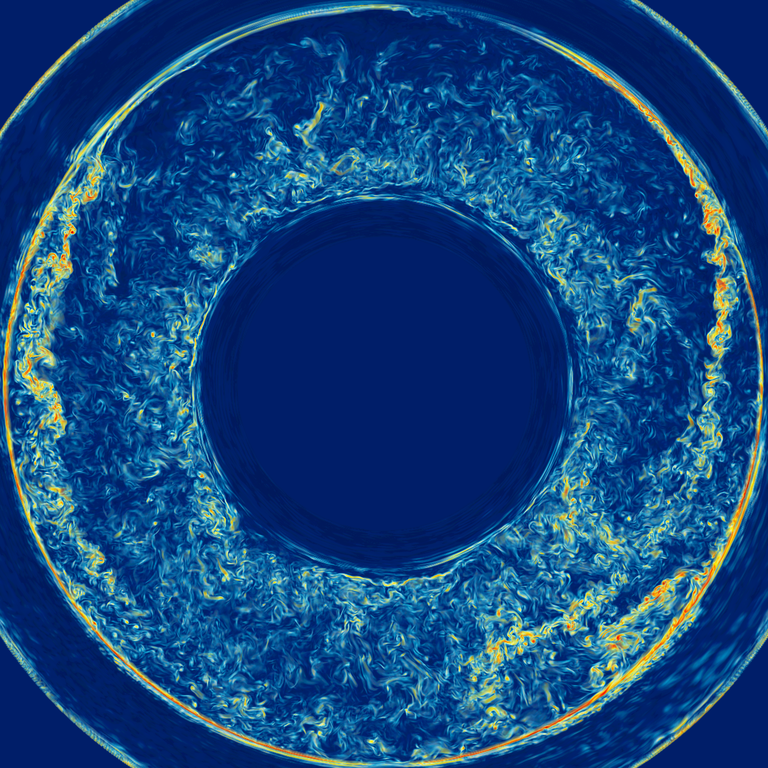}
\includegraphics[width=0.333\textwidth]{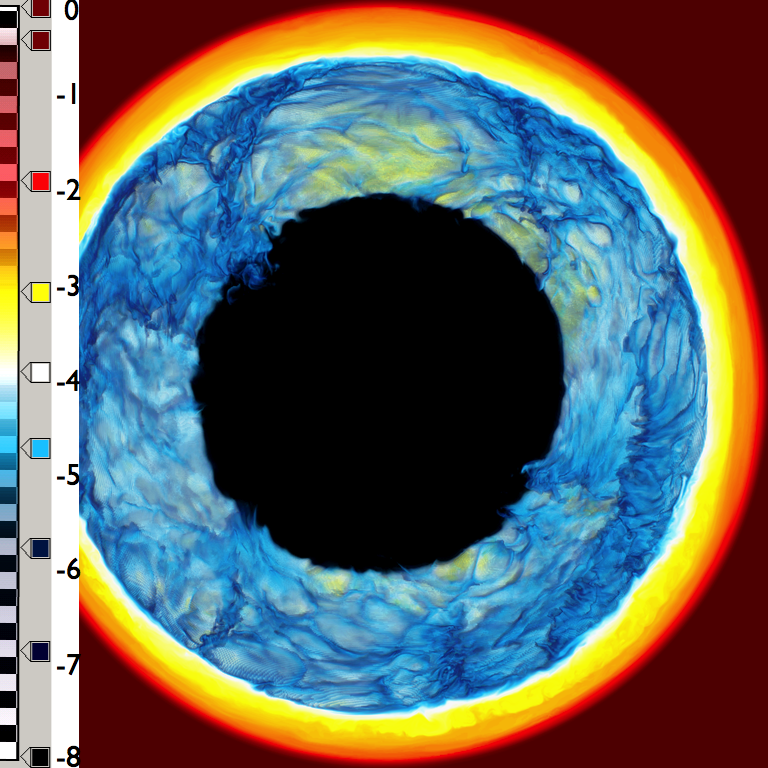}
\includegraphics[width=0.333\textwidth]{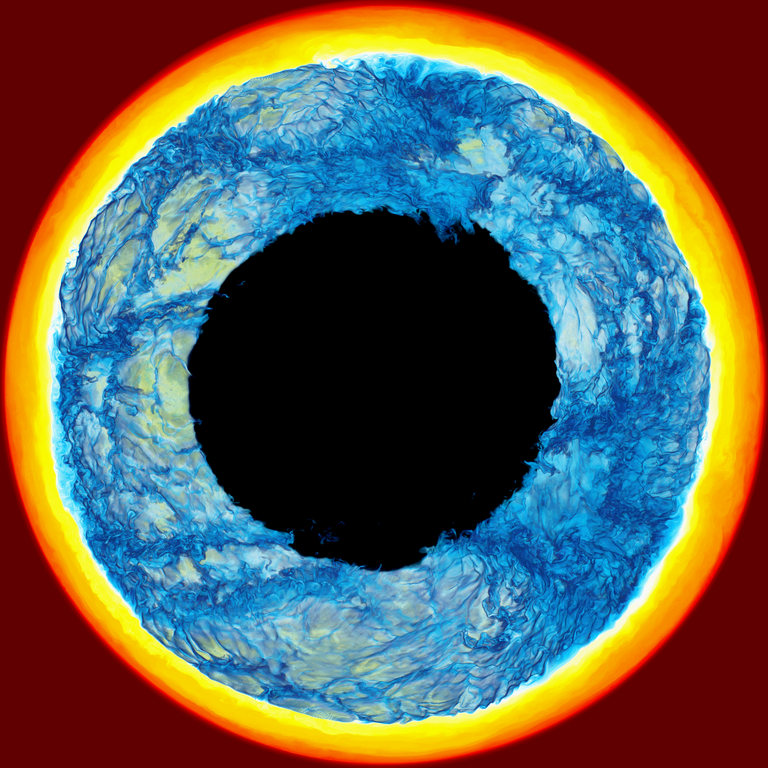}
\includegraphics[width=0.333\textwidth]{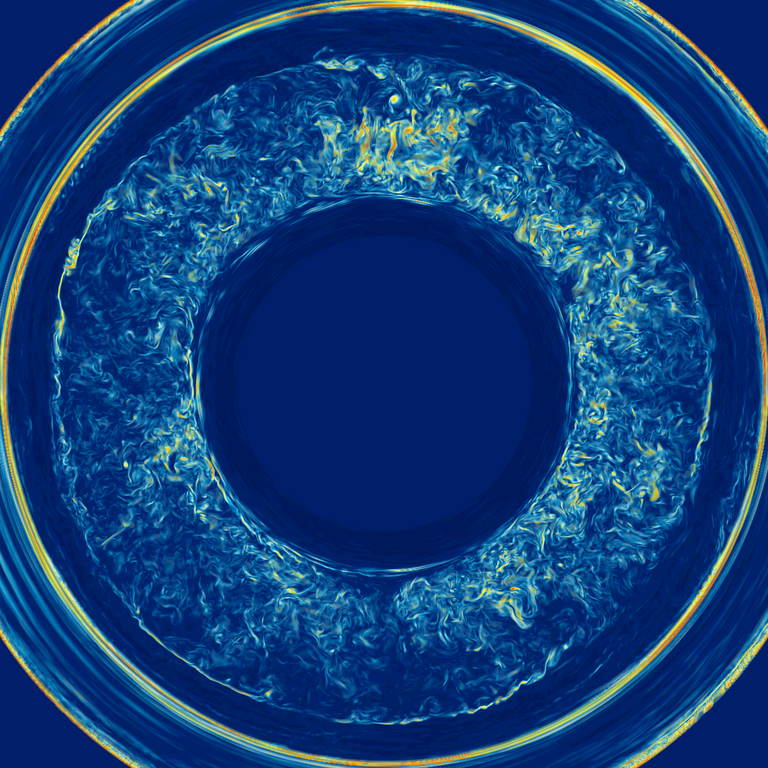}
\caption{Fractional volume of fluid H+He from $768^3$ (left) and
  $1536^3$ (middle) grid runs and vorticity (right) from $1536^3$ run
  at $960\mathrm{min}$ (top row) and $1100\mathrm{min}$ (bottom
  row). The colorbar indicates mapping of $\log$ of fractional volume
  of 'H+He' fluid as well as the transparency used for the volume
  rendering.  In each panel a slice with thickness equal to one
  quarter of the domain is shown, with the near surface going through
  the center of the star.  In the viewing plane the horizontal and
  vertical limits are at the position where the upper convection
  boundary is initially located.} \label{fig:FV_vort}
\end{figure*}

\begin{figure}
\includegraphics[width=0.5\textwidth]{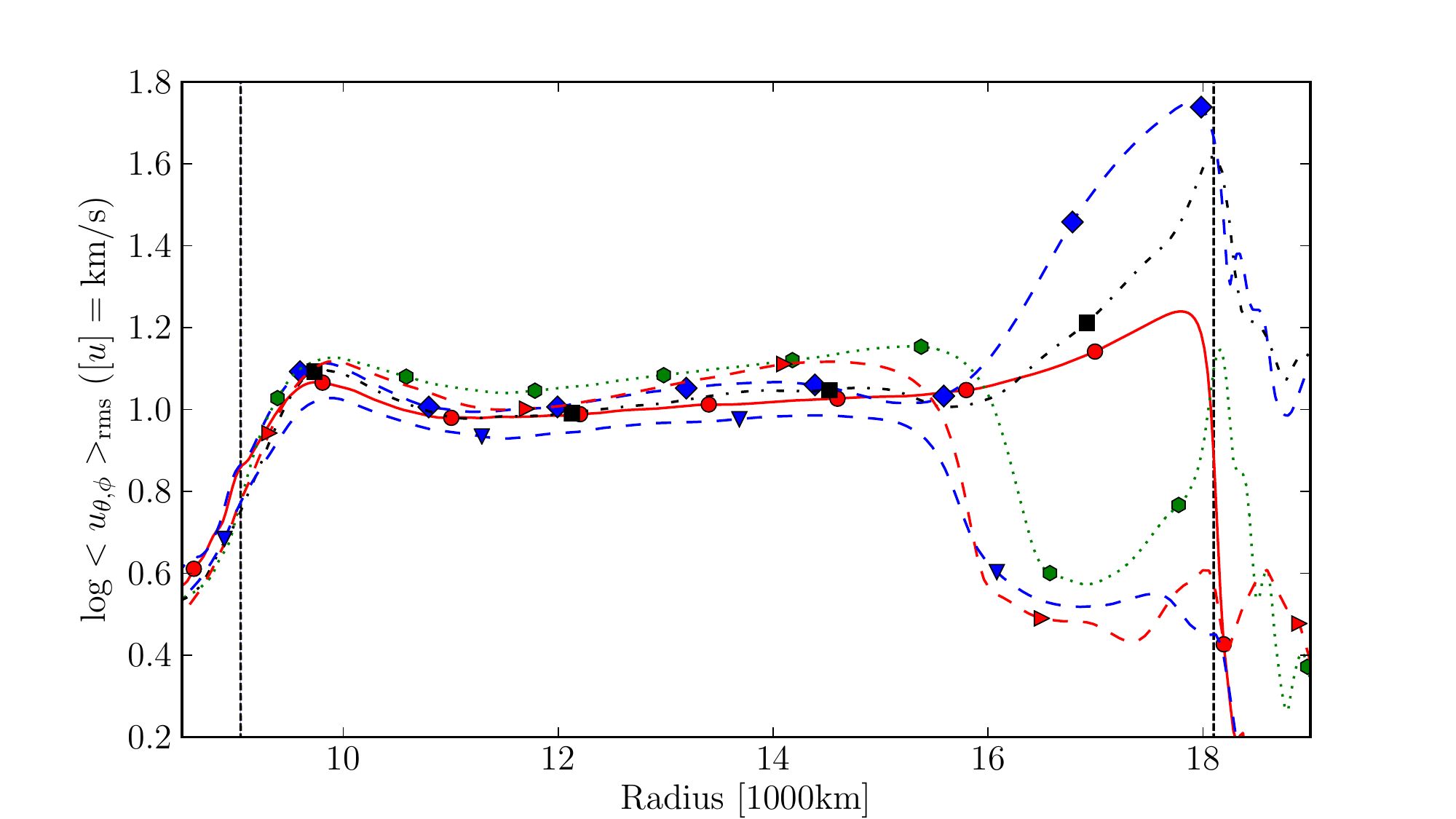}
\includegraphics[width=0.5\textwidth]{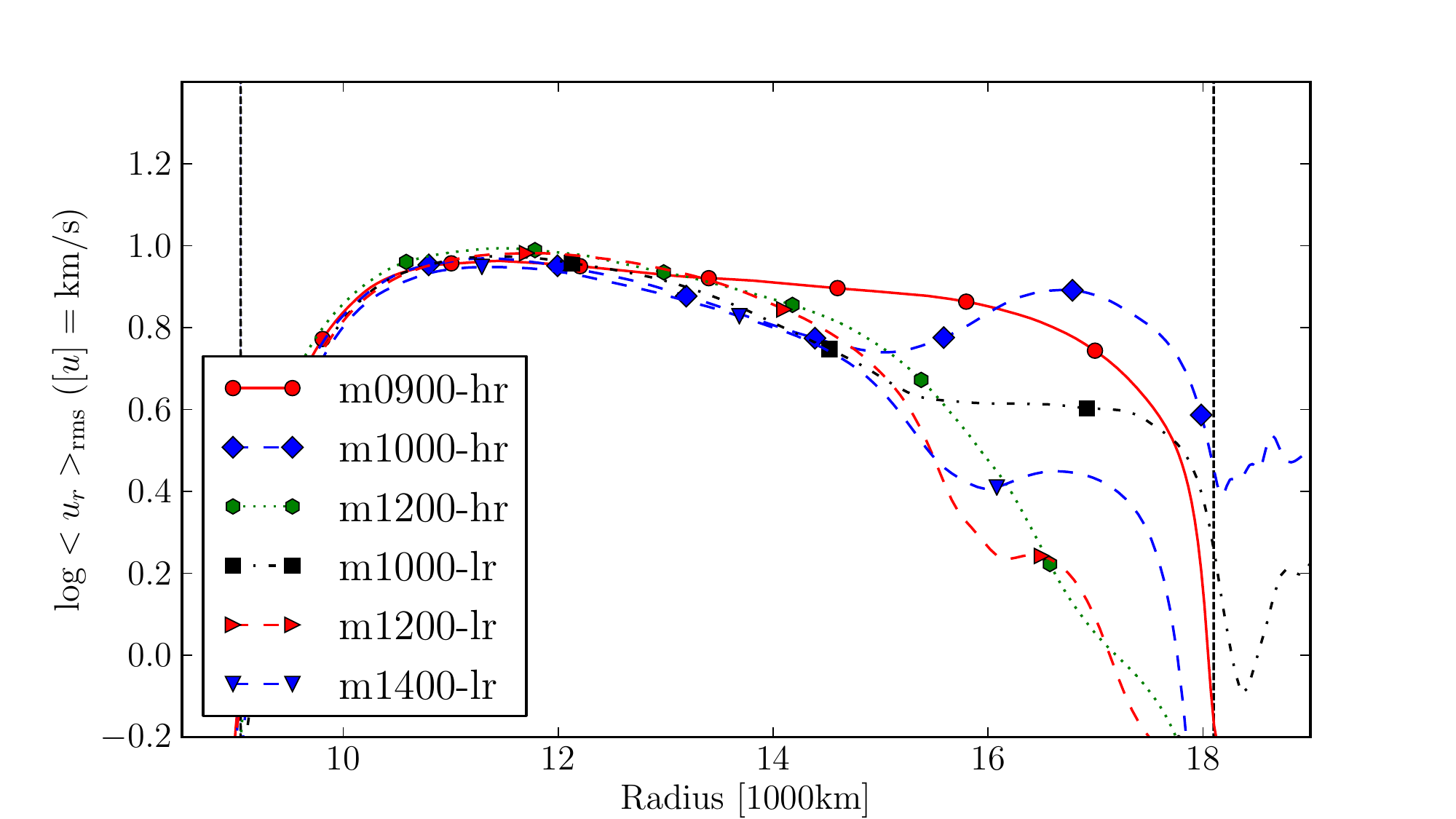}
\includegraphics[width=0.5\textwidth]{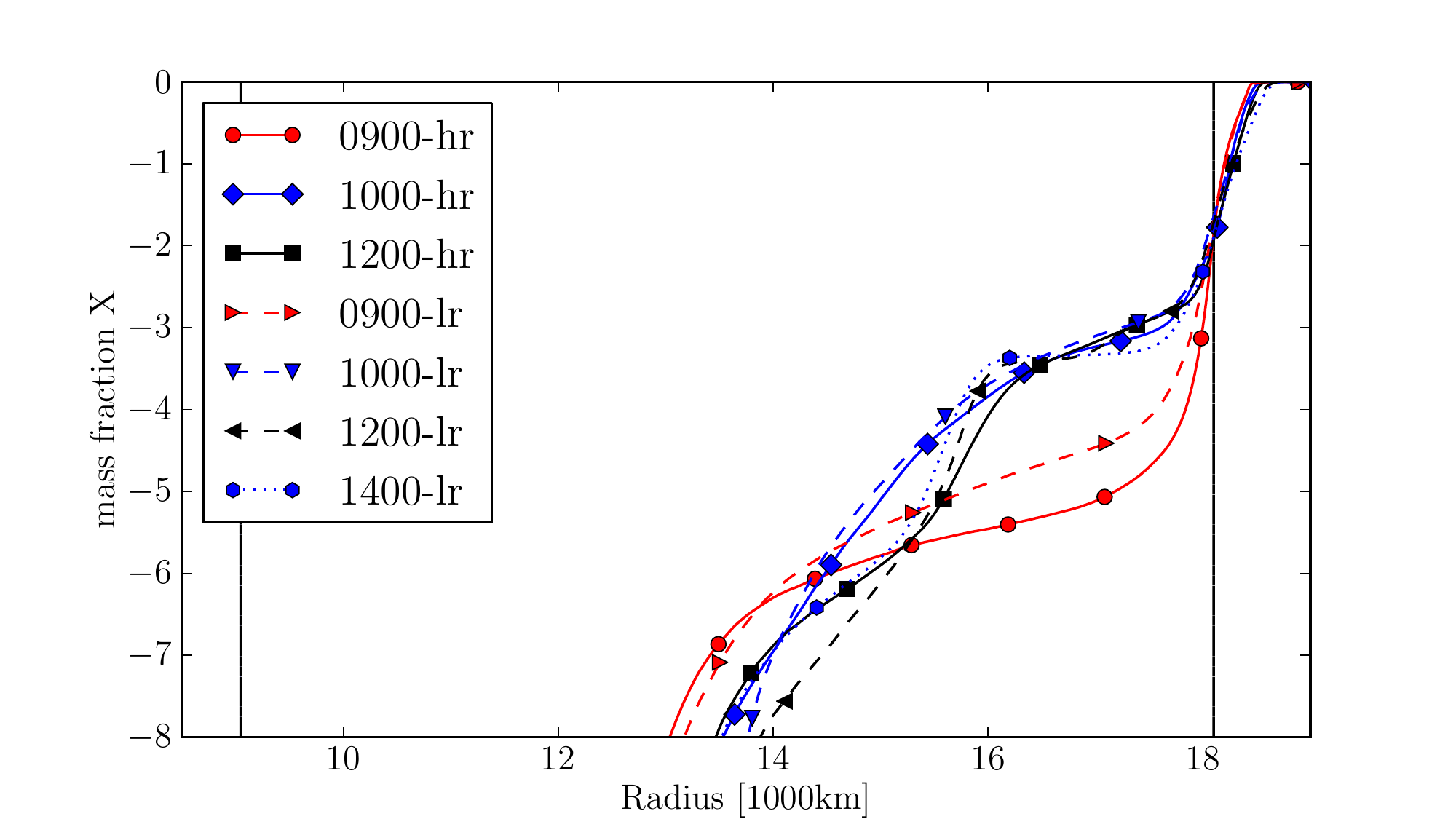}
\includegraphics[width=0.5\textwidth]{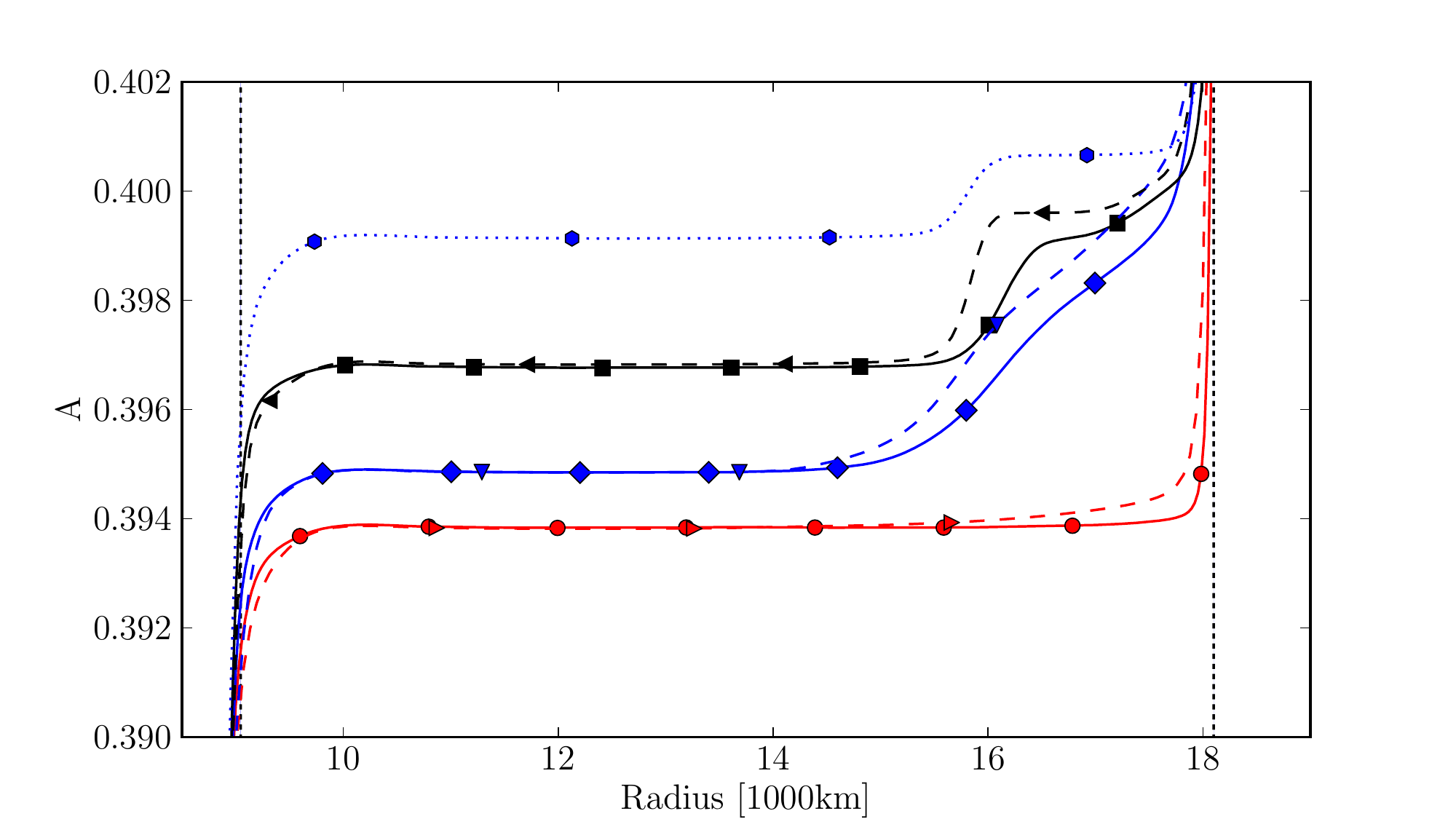}
\caption{Top and middle panel (same line styles): tangential and
  radial spherically averaged rms velocity components for different
  times (labels  indicate time in
  minutes) for the $768^3$  (lr) and the $1536^3$ (hr) runs.
  $3^\mathrm{rd}$ panel: Abundance profiles of H-rich
  material. Bottom panel: entropy-like quantity $ A = p/\rho^\gamma $,
  with $p$ pressure and $\gamma = 5/3$ for the monatomic ideal
  gas. The bottom two panels have the same line styles. Vertical
  dotted lines indicate the convection boundary in the initial
  setup.} \label{fig:profiles}
\end{figure}

These examples demonstrate the challenge that these phases of stellar
evolution present to simulation efforts. It would be very helpful to
check the simulation approach for a case for which meaningful
validation data is available to check the results.  Therefore, before
turning our attention to the interesting problem of H-ingestion events
in metal-poor stars, we perform for the first time 3-D hydrodynamic
simulations of the H-ingestion flash in Sakurai's object (V4334 Sgr),
a post-AGB star that experienced a very late thermal pulse
(VLTP). Real-time observations of the light curve \citep[e.g.][]{duerbeck:96,hajduk:05} as well as post-flash
abundance observations \citep{asplund:99a} provide constraints on the
details of the He-shell flash with H-ingestion in this case.

\citet{Herwig:2001km} showed that the $\approx
100$ times faster rise time of Sakurai's object compared to standard
mixing-length theory (MLT) stellar evolution predictions can be
accounted for if it is assumed that in H-ingestion phases the
convective mixing efficiency is reduced by a factor of $30$ to $100$,
which leads to a H-combustion energy release closer to the stellar
surface, where the time scale for thermal response is shorter.

\citet{herwig:10a}
showed that an early mixing split between the He-shell flash
convection zone (reduced in size) and the newly emerging H-burning
convection zone prevents standard stellar evolution models from
reproducing the observed abundance features, especially the pronounced
overabundance of first-peak neutron capture elements. However, an
assumed mixing permeability of the interface between the two
convection zones lasting for $\approx 900 \mem{min}$ after the
H-ingestion event started does enable one to account for most observed
abundance features.

 In both cases 1-D stellar evolution assumptions fail to reproduce key
 observables. In \kap{sec:method} we
 describe the method and how the simulations were performed, while \kap{sec:results}
 presents our main findings and discussion of the results.

\section{Methods and simulations} 
\label{sec:method} 
We use the 3-D gas dynamics code and the initial setup method described
in detail in \citet{Woodward:2013ui}.  The code features the
Piecewise-Parabolic-Boltzmann (PPB) moment-conserving advection scheme
\citep[see][]{woodward:08a}\nocite{woodward:08b}, which provides an
effective resolving power of the fluid distribution that requires two
to three times the grid resolution to be matched by a code based on
the Piecewise-Parabolic Method (PPM) without PPB.  As in previous
investigations \citep[e.g.][]{porter:00a,herwig:06a} we adopt a
monatomic ideal gas equation of state which represents the conditions
in He-shell flash convection in the post-AGB He-shell flash model for
Sakurai's object well (\abb{fig:initial_star}).

The nuclear burn module considers just the $\czw(\p,\gamma)\ndr$
reaction \citep{angulo:99} with an energy release  of
$Q=1.944\mathrm{MeV}$. The two-fluid setup consists of a convectively
unstable shell and a stable layer below with initially the \czw-rich
fluid 'conv', and a stable layer above the convection zone with the
H-rich fluid 'H+He'. Both of these composite fluids have realistic
mean molecular weights reflecting the abundance mix for intershell and
envelope material according to the corresponding stellar evolution
model. The nuclear burn module takes appropriately into account that
in each of the fluids only a certain fraction participates in the
nuclear reaction. We ignore for now the energy release of
$1.508\mathrm{MeV}$ from the subsequent $\beta$-decay of \ndr\ with a
half life of $9.96\mathrm{m}$. Using the reaction rate length scale
introduced by \citet{herwig:10a} as a 1-D estimate of the geometric
scale of the H-burning layer \ndr\ would decay after being homogenized
throughout a layer approximately 20 times the thickness of the burn
layer. Here we are interested in the emergence and initial evolution
of any hydrodynamic response to H-ingestion and burning which we
assume to depend on the localized energy release from the $\czw+\p$
reaction.

The full $4\pi$ shell of the convectively unstable layer is included
in the simulation, which ensures that any global or large-scale
motions can be captured.  The initial setup has been constructed by
using the model at $t=t_\mem{0}$ from \citet{herwig:10a}, which was
taken from the $0.604\msun$ post-AGB stellar evolution track presented
first in \citet{Herwig:1999uf} and further modified in
\citet{Herwig:2001km}. The convection is driven by a constant heating
of $\natlog{4.75}{7}\lsun$ corresponding to the He-shell flash
luminosity of the $t=t_\mem{0}$ stellar evolution model. The initial
stratification (\abb{fig:initial_star}) is constructed from piecewise
polytropic layers that closely resemble the stellar evolution model. A
reflective spherical outer boundary condition is located at
$\natlog{22}{3}\mathrm{km}$.

For our simulation approach we have investigated in detail the
numerical resolution requirements \citep{Woodward:2013ui}. In addition
we have performed numerous test simulations for the Sakurai's star
setup, on the Canadian WestGrid computer Orcinus with grid sizes up to
$768^3$ while simulations with $768^3$, $1152^3$ and $1536^3$ grids
where performed on NSF's Blue Waters sustained petaflops computing
system at NCSA.

The PPMstar code \citep{Woodward:2013ui}, which was originally
designed to run on the Los Alamos Roadrunner machine
\citep[cf.][]{Woodward:2009hl}, performed our run at $1536^3$
resolution within a single four-day period on Blue Waters.  It ran on
$443,232$ CPU cores, roughly half the Blue Waters machine, at a
sustained rate of $0.42 \mathrm{Pflop/s}$ with 64-bit precision.  At
this computing rate, it took  about $3\mem{min}$ of real time to
simulate $1\mem{min}$ of time for the star and to write to disk
$41.5\mem{GB}$ of visualization data.  With $1362$ data dumps,
$56.5\mem{TB}$ of data from this run was written, which was used to
generate the movies at \url{http://www.lcse.umn.edu/movies}.  The
movies were generated using our HVR volume rendering utility on the
GPU nodes on Blue Waters.
This computational performance was possible, because our code scales
extremely well on Blue Waters.  Each node with $32$ CPU cores ran
eight MPI processes, each with four threads updating a tiny grid brick
of just $32^3$ cells.

\section{Results and discussion}
\label{sec:results}

We present the results from two runs with a $768^3$ a $1536^3$ grid
for a minimum duration of $20\mem{hr}$ star time.  The simulations
start in the same way as the entrainment simulations presented by
\citet{Woodward:2013ui}. An initial transient phase of less than
$50\mathrm{min}$ quickly gives way to fully developed He-shell flash
convection with dominant global modes that cause hydrodynamic shear
instabilities at the top convection boundary. Downdrafts of H-rich
material from above originate where large-scale upwellings and
consequently coherent horizontal motions along the convective boundary
converge. H-rich clouds descend into the deeper layers and represent a
large-scale inhomogeneous distribution in the region where protons
start to react with \czw.  During this initial phase of $\approx
850\mathrm{min}$ the spherically and radially averaged velocities in
both runs agree reasonably well (\abb{fig:initial_star}, bottom
panel). The overall entropy in the convection zone rises due to the
addition of energy from nuclear reactions, and mixing throughout the
convection zone is efficient enough to maintain a flat spherically
averaged entropy profile. The H-accumulation rate (entrainment minus
burning) is $1.81$ and $\natlog{1.04}{-13}\msun/\mathrm{s}$ for the
$768^3$ and $1536^3$ runs during this initial quiet phase, about a
factor four smaller than the converged entrainment rate of
$\natlog{4.38}{-13}\msun/\mathrm{s}$ found by \citet{Woodward:2013ui}
for slightly different runs without H burning. As discussed there,
this entrainment rate is too small to cause hydrodynamic feedback from
the burning of entrained H if the entrainment rate is equated with the
H burn rate. It is therefore consistent that we do not observe
noticeable differences in the early quiescent phase between these
simulations with H burning and the entrainment simulations of
\citet{Woodward:2013ui} without H burning. The quiescent phase lasts
for around $850\mathrm{min}$.  In test simulations with a sponging
boundary condition the motions in the upper stable layer are much more
pronounced in this setup. These and other details of the simulations
can alter the duration of the quiescent phase significantly. However,
none of our tests indicate that nature of the violent events to follow
depend on the details of the quiescent lead-up phase.

The convection hydrodynamics changes dramatically around $850\mathrm{min}$
(\abb{fig:initial_star}) when a major global fluid flow oscillation is
about to be launched. The overall shape of the rise and decay of the
oscillation mode agrees well between the two runs, the major
difference being that in the lower resolution run the event starts
$20\mem{min}$ earlier.  The visualization of the 3-D flow for time
$960\mathrm{min}$ shows the moment when the flow has already started
to respond strongly to the energy input from H burning
(\abb{fig:FV_vort}). The
fractional volume images from the two runs appear to differ more than
at the later time ($1100\mathrm{min}$). However, this is mostly because we see in the
$768^3$-run image a later time in the event. Overall the agreement
between the two runs is good.

During the lead-up to this transition at $\approx 850\mathrm{min}$
eventually a large enough reservoir of H is accumulated in the upper
part of the convection zone (\abb{fig:profiles}) so that pockets of
H-rich material advected into the burn region carries an increasing
fraction of fuel. At that point individual, particularly strong
ignition events launch very pronounced, identifiable upwellings. As
these hit the stiff convective boundary above they are deflected in
all directions and create enhanced levels of entrainment. The
Kelvin-Helmholtz instability vorticity-entrainment trains can be
clearly identified in the vorticity image (top row,
\abb{fig:FV_vort}). The horizontal flows proceed along the convective
boundary all the way to the antipode where their collision is forcing
the H-rich entrainment layer into a downdraft. This next fuel-enhanced
advection stream causes an even more violent ignition event and
subsequent launching of the next fast upwelling, which in turn causes
the next circumferential flow back close to the point of origin of the
first ignition. This oscillatory back-and-forth repeats for about a
dozen times and is clearly identifiable in the averaged velocity time
evolution shown in the bottom panel of \abb{fig:initial_star}. We
refer to this new phenomenon in stellar physics as the Global
Oscillation of Shell H-ingestion (GOSH). During the rise time of
$\approx 100\mathrm{min}$ $4.9$ and $\natlog{4.2}{-7}\msun$ are
accumulated in the $768^3$ and $1536^3$ runs, corresponding to an
accumulation rate of $\natlog{7}{-11}\msun/\mathrm{s}$ and close to
the critical entrainment rate for hydrodynamic feedback estimated in
\citet{Woodward:2013ui}. These highly time-variable and global fluid
flows are better observed in the animated movies available at
\url{http://www.lcse.umn.edu/movies}.

Although the GOSH proceeds beyond the initial rise time, its amplitude
decreases until convective motions in the upper region of the
initially unstable zone are significantly suppressed (see vorticity
image at later time and spherically averaged radial and horizontal
velocities shown in \abb{fig:profiles}). This suppression follows from
the entrainment of entropy that goes along with the entrainment of
H-rich material from the stable layer above the convection zone (lower
panel \abb{fig:profiles}). In the simulation a new upper boundary of
the He-luminosity driven convection forms at
$\natlog{15.5}{3}\mathrm{km}$ that can be seen clearly in both runs
for the later time in the fractional volume and the vorticity
image. However, the burning of entrained H-rich material continues to
add entropy to the convection zone as well and causes an entropy
plateau to develop in the upper region. The $1536^3$ run did not yet
get to that point when we stopped it. But the $786^3$ run with its
generally higher entrainment rate shows an upper convection zone is
about to form between $1200\mem{min}$  and
  $1400\mem{min}$.

  When we continued this lower-resolution run beyond
  $1400\mathrm{min}$ it developed a further non-radial, violent event
  in which the velocity amplitude in the upper layer above the initial
  convection zone increases significantly. This simulation outcome is
  not reliable because the outer boundary of the domain needs to be
  moved further out to simulate this following phase accurately.
  During the GOSH significant and mostly tangential motions are set in
  place in the stable layer above the convection zone that die down
  when the oscillations subside.  In terms of the averaged velocities
  (as those shown in \abb{fig:profiles}) these motions are agreeing
  fairly well between the two grid simulations up to
  $1100\mathrm{min}$ when the GOSH is ending. At later times larger
  differences are evident. This is the phase that we need to explore
  in more detail with higher resolution grids.

What connections can we make between these preliminary results of the
3-D hydrodynamic simulations and the constraints we have so far from
comparing simulations with calibrated 1-D
models? \citet{Herwig:2011dj} reproduced observed n-capture element
abundances by assuming a constant entrainment rate of
$\natlog{5.3}{-10}\msun/\mem{s}$ for $900$ to $1000 \mem{min}$ before
terminating further neutron exposure by assuming mixing between the
upper and lower region of the convection became inefficient. Our
simulations so far show that compared to the very low entrainment we
found in He-shell flash convection without feedback from H burning
\citep{Woodward:2013ui}, entrainment rates are significantly enhanced
by the GOSH. However, the GOSH leads as well, at least initially, to a
self-quenching of convective motions close to the original upper
convection boundary due to the entrainment of entropy. Convection is
restarted in this layer by H burning and during one (or perhaps several)
subsequent violent outburst(s) the entrainment rates may further
increase. These enhanced entrainment rates will have to be sustained
so that the observed 2-dex enhancement of first-peak elements Sr, Rb,
Zr, Y can be reproduced. In which way this may be possible will have
to be revealed by future simulations. The present simulations show
that the H-ingestion is proceeding through periodic bursts of
global-mode oscillations and highly variable entrainment rates, and we
will have to investigate if such entrainment histories would
fundamentally allow the reproduction of the observed abundance
features in Sakurai's object.  In our present simulations the
reflective boundary condition is likely unrealistic when the flow
becomes very violent in the initially stable layer above.

The location of H-ingestion energy release is related to the
time-scale of the born-again light curve of VLTP objects
\citep{Herwig:2001km}. During the long quiescent phase of our 3-D
simulations H burning takes place in deeper layers of the He-shell
flash convection zone, around $\natlog{13}{3}\mem{km}$. However, when
the feedback from H burning becomes evident burning moves up to
$\approx 15$ to $\natlog{16}{3}\mem{km}$, and here the new convective
boundary forms after the GOSH.  \citet{Herwig:2001km} has associated
the light curve time scale of Sakurai's object with the mixing
efficiency. Our 3-D setup (\abb{fig:initial_star}) is based on this
same $0.604\msun$ 1-D model used by \citet{Herwig:2001km}. The
unmodified stellar evolution track shows the H-burning location at
$m_\mathrm{r}=0.595\msun$ (corresponding to a radius
$r=\natlog{12.0}{3}\mathrm{km}$) with a born-again time of $>
200\jahre$, compared to $m_\mathrm{r}=0.601\msun$
($r=\natlog{15.5}{3}\mathrm{km}$) for a modified model with reduced
mixing efficiency calibrated to reproduce the observed time of
$2-3\jahre$. The entropy profile (\abb{fig:profiles}) shows that the
H-burning location in the 3-D hydrodynamic simulation indeed agrees
very well with the lower-mixing efficiency solution of the calibrated
1-D model, and would lead to a fast born-again time as observed. This
agreement between the location of H-burning in a stellar evolution
model calibrated from the light curve, and the location found in the
3-D simulation that has no free parameters provides a first validation
of our simulation approach.

Our simulations show highly non-radial, global and variable modes with
periods of the order of a convective turn-over time. These global
modes of the H-ingestion event cannot be captured in 1-D
models. However, 1-D stellar evolution simulations do agree
with some aspects of our 3-D simulations, as for example the
relatively early emergence of a H-burning convection zone in terms of
the entrained mass of H, and the fact that energy-release from
H-burning has some influence on the convective properties of the
He-burning shell. This is, for example, opposite to the finding of
\citet{stancliffe:11} that H is mixed all the way to the bottom of the
He-shell flash convection zone (their Fig.\,9), a result that we do
not confirm in our 3-D simulations (\abb{fig:profiles}). However, the
low-Z AGB thermal pulse has due to lower density a smaller gas
pressure fraction ($\beta \approx 0.8$) compared to post-AGB
H-ingestion case where $\beta \approx 0.96$.  This may limit direct
comparisons between their and our work. Another difference is that the
transition to a H-burning driven convection zone that we report here
appears after $\approx 54000\mathrm{s}$ in our simulation, whereas the
\citet{stancliffe:11} runs stop after $14400\mathrm{s}$. Perhaps those
authors did not see a significant hydrodynamic response from H-burning
energy release because they stopped the run too early. However, as we
have pointed out here, even with our much longer simulations we can be
certain that we have not yet seen the full story of the H-ingestion
flash.

Finally, we note that our results reported here have to be considered
as preliminary.  We have to postpone a more detailed analysis of the
results of our simulations until we are able to carry the high
resolution run further forward in time.

\acknowledgements The computer simulations have been performed in the
US on the NCSA BlueWaters system under PRAC grant NSF/OCI-0832618 and
in Canada on the WestGrid Orcinus computer. FH acknowledges funding
from an NSERC Discovery Grant.  PRW acknowledges DoE support from
contracts with the Los Alamos and Sandia National Laboratories, NSF
CRI grant CNS-0708822, and support from an NSF subcontract from the
Blue Waters project at NCSA.


\end{document}